# Conversational agents for learning foreign languages a survey

Jasna Petrović [1*], Mlađan Jovanović [1]

[1] Singidunum University, Belgrade, Serbia
[*] jpetrovic@singidunum.ac.rs (corresponding author)

*Abstract* - Conversational practice, while crucial for all language learners, can be challenging to get enough of and very expensive. Chatbots are computer programs developed to engage in conversations with humans. They are designed as software avatars with limited, but growing conversational capability. The most natural and potentially powerful application of chatbots is in line with their fundamental nature - language practice. However, their role and outcomes within (in)formal language learning are currently tangential at best. Existing research in the area has generally focused on chatbots' comprehensibility and the motivation they inspire in their users. In this paper, we provide an overview of the chatbots for learning languages, critically analyze existing approaches, and discuss the major challenges for future work.

## I. INTRODUCTION

Conversational agents (also known as chatbots) are becoming part of our everyday activities, such as Amazon Alexa [1] and Google Assistant [2]. These services employ a common technological metaphor - a personal, user-friendly assistant that provides a service by using a natural language [3].

The proliferation of technologies for learning languages aims at providing personalized services for gaining and maintaining language proficiency at any time, in any place [4]. Chatbots are becoming part of this paradigm shift as a cost-effective means to deliver such services. Besides, they try to ingrain positive working habits concerning language, such as repetition and follow-up exercises outside the classroom [5], [6]. The main benefits are ease of use and accessibility - the conversation metaphor using text- or voice-based interfaces make them more intuitive, available on smartphones in any place and at any time [7].

Recent research with various text-based human-chatbot interactions has consistently pointed to their potential benefits. Particularly concerning the motivation and engagement they inspire in their users [6], [7]. These are critical requirements for technology-supported language learning. However, the potential of chatbots as language learning partners is yet underexplored. The evidence of the effectiveness and efficiency of the existing chatbot-like technologies for learning languages in terms of measurable outcomes (such as acquired knowledge, skills, and attitudes) is either missing or is non-consistent [4], [5], [6], [7], [8], [9].

This situation motivates our research to reflect on the previous work in chatbots for learning languages and formulate implications for their future. Due to their very nature, we believe that the chatbots' appearance and behavior are highly interdependent. Therefore, we look at them from technological (i.e., functional) and speaker (i.e., user experience - UX) aspects. In particular, this paper:

- gives an overview of the technology for learning languages,
- critically analyzes the chatbot technology against the proposed dimensions, and
- outlines the implications for future work on the chatbots for learning languages.

The paper is structured as follows: Section 2 situates our work in the context of the related technologies. In Section 3, we formulate the dimensions for analyzing relevant aspects of the chatbots for learning languages. Section 4 describes and compares the chatbots against the dimensions. Section 5 articulates the implications for future work. Section 6 proposes a conceptual architecture for the chatbots. Section 7 highlights the limitations of our work, and Section 8 concludes the paper.

## II. TECHNOLOGY FOR LEARNING LANGUAGES

Numerous applications for learning languages have been developed and used. We classify these applications into Web and mobile technologies and gamified technologies based on their underlying mechanisms and type of content delivery. Given their abundance, we do not aim to provide an exhaustive list of the technologies but analyze the representatives that demonstrate some shared properties.

*Web and mobile technologies*

A lack of time, numerous commitments and scheduling issues among others have led to people having to find ways around to learn foreign languages to maintain regular contact with the target language and practice to become more fluent and proficient. In doing so, they seek assistance from diverse language learning programs that can be used on desktop computers, tablets, or smartphones and are always available to the user.

For this study, the authors have tested several popular language learning applications. HelloTalk [10] is a mobile application whose users learn foreign languages through a conversation with native speakers. The users can search for and chat with learning partners (autonomous learning), or join online group conversations moderated by teachers (collaborative learning). Duolingo [11] follows a more systemic and scaffolded approach in that it targets specific language skills (such as speaking, reading, listening, and writing skills) through a series of exercises. Rosetta Stone [12] is a universal platform for learning languages. To attract users, it implements different learning strategies. The strategies include topic-based learning (such as work, home, and similar), immersive experiences from rich multimedia content with high interactivity, skill-specific lessons (i.e., vocabulary, grammar, and others), or interactive storytelling on various topics. Memrise [13] facilitates language acquisition through learning by example (users can watch videos of native speakers conversing about the object they just saw or that came in their mind), learning by exploration (object identification in real-time so that a user can make a photo of an object and ask for a description in the chosen language from native speakers), spaced repetitions, or instant feedback on pronunciation. WordUp [14] is an application for learning English vocabulary. It aims to increase and maintain the user's vocabulary by prioritizing words based on their usefulness, putting words in familiar contexts, and repetition.

We explored English language learning opportunities. Given that the language of instruction could not be Serbian, we had to rely on our knowledge of either Russian, Croatian or Italian (as there is no English language instruction option for learning the English language). To gain a better understanding, especially when it comes to motivation to learn a foreign language, the authors of this paper, in addition to English, tested learning French and Italian at A1 level and Russian at B1 level.

All applications share some common features. Each application contains a brief grammar and/or vocabulary tutorial that serves as a preparation for or a reminder after the lesson has been delivered. The introductory test available in every application determines the user's language level, and following the results obtained, offer a course tailored to the learner's needs. However, it is unclear how the level of language proficiency is assessed, since for the fluent speaker, the tasks come across as too easy, which in return may result in a quick loss of interest. We found that all the applications are largely intended for those whose level of knowledge corresponds to the basic or elementary level, and that for any advanced progress, relying on this kind of instruction is not sufficient. The tasks generally consist of translating from one language into another, listening and spelling out what is heard (clicking on the words offered), inserting verbs or nouns or articles into the appropriate form, and in the appropriate place with many repetitions. As the user progresses, the application adapts its content and functions to the users' language skills.

Maintaining continuity is certainly what every application insists on. In doing so, the applications employ the motivational elements of self-reflection (the users are continuously monitored and presented with their achievement), competition (comparison with peers), and digital nudging (regular reminders for the exercises). Except for WordUp, all applications support multiple languages. Most apps are free to some extent and charge their users for more advanced functions.

*Gamified technologies*

The term gamification is defined as using game elements in non-gaming contexts [15], such as learning foreign languages. These applications use elements of games to incentivize their users, where learning happens as a side effect of playing the game. The user incentives are internal and include fun, pleasure, satisfaction, challenge, achievement, reputation, and competition, among others [16], [17]. For example, Duolingo [11] employs aesthetical visual metaphors of hearts, badges, winning streaks, and leaderboards to motivate their users [18]. The Kahoot! [19] is an online educational platform used for gamified language learning [16]. The gamified components are multiple-choice and trivia quizzes that can be created and moderated by teachers. Vertaal [20] follows a similar approach while using memory games and crosswords to facilitate learning foreign languages [16].

The study described in [17] provides a systematic review of digital games that follow a scaffolding approach to vocabulary acquisition of foreign languages. The games' genres vary and include storytelling games, problem-solving games, role-based games, virtual reality (VR) and body-motion games, adventure games, card games, and board games. While there is partial evidence of the games' positive effects on vocabulary acquisition, the issues of more sustained user engagement and a broader outreach remain. The issues are due to the inherent diversity of users' motivations to play games, and the heterogeneity of user groups in terms of needs, preferences, and learning styles.

A similar analysis conducted in [16] shows that the digital game-based language learning induces positive emotions and attitudes of students. However, the games' positive effects are observed in a controlled environment (classroom). Furthermore, the games are biased towards the English language and mainly improve the vocabulary while having less influence on other language skills (such as reading, writing, speaking, and listening).

III. DIMENSIONS FOR ANALYSIS

We propose the dimensions to describe and compare existing chatbots for learning foreign languages (Table 1). The dimensions are extracted as themes from an emerging, inductive coding process on the existing solutions. The chatbots were obtained by an extensive search conducted by the authors from the available resources, including chatbots' websites (shown in Table 2), public collections and articles [21], [22], [23], [24], [25], and peer-review publications [4], [5], [6], [7], [8], [9], [26].

The dimensions are classified into *speaker* and *technology* aspects.

The *speaker aspects* include skill level, language diversity, domain, and interaction modality. The skill level tells whether the chatbot can adapt to the speaker concerning general language proficiency at a single or multiple levels of knowledge and skills. The language dimension indicates whether the chatbot supports learning one or more languages. The domain shows whether different thematic areas of knowledge are offered within a single language. The modality can be a text-based conversation (possibly including visual elements), speech-based conversation, or multimodal (combining the previous two).

Concerning the *technology*, the dimensions include dialog type, knowledge base, and availability. As for the dialog type, we adopted the existing taxonomy of chatbot dialog management systems as 1) predefined, based on the matching of a finite set of rules, and 2) statistical that generates responses dynamically from a larger dataset employing machine learning (ML) algorithms [27]. The knowledge base indicates the openness of the chatbot's data source of the linguistic concepts and terms for the language(s) it supports. Finally, the availability shows whether the chatbot is free to use or charges users for its services (fully or partially).

TABLE I. ANALYTICAL DIMENSIONS.

| Aspect | Dimension | Value | Description |
|---|---|---|---|
| Speaker | Skill level (SL) | single | Single or not explicitly separated levels |
| | | multiple | Multiple levels explicit for user |
| | Language (L) | single | Single target language |
| | | multiple | Multiple target languages |
| | Domain (D) | single | Single knowledge domain |
| | | multiple | Multiple knowledge domains |
| | Modality (M) | text | Textual communication (with possible visuals) |
| | | voice | Speech communication |
| | | multimodal | Multimodal communication (e.g., text, voice, media) |
| Technology | Dialog type (DT) | scripted | Predefined, rule-based dialog management |
| | | statistical | Flexible, ML-based dialog management |
| | Knowledge base (KB) | open | Open, known language knowledge base |
| | | closed | Closed, proprietary language knowledge base |
| | Availability (A) | open | Free service |
| | | commercial | Paid service |

IV. CHATBOTS FOR LEARNING LANGUAGES

The authors independently installed and probed the chatbots against the dimensions as of May 2020. The individual analyses were compared and reconciled in a series of discussions to reach an agreement on the chatbot's analytical evaluation shown in Table 2.

Looking at the table, we notice that the current chatbots mainly: offer multiple skill levels and languages to learn, combine knowledge from more domains, provide textual conversations, implement scripted user dialogs using proprietary linguistic knowledge bases, and charge their users.

Before we describe the examples, how they work, what functions they possess, their benefits and downsides, it's worth mentioning that even though the reference list of chatbots is longer, the number of real (proper) or active chatbots is much smaller. Namely, as seen in the table below, we have tested only 4 conversational agents that support real-time dialogue with users, providing immediate answers to prompts, and asking relevant follow-up questions. The currated chatbots [21], [22] including Edwin, Instant Translator, Translator ChatBot, Grammar Guru, CallMom, Tutor Mike, and Dave English Teacher were not available to test.

**Mondly** is a language learning application which is one of the few that, in addition to modern and interactive ways of learning languages through daily lessons that cover a plethora of topics (family, travel, restaurant to name a few) at the same time offers a voice-enabled chatbot. The application itself is partly free of charge, and in order to get access to all the chatbot's features - the user must subscribe to the content to unlock all the lessons. We only had a glimpse of the first lesson with instructions in the Croatian language - the lesson is called Hi, and the user is at liberty to go back to the same lesson over and over again. The communication starts with a greeting (e.g., "Good morning") and the user may select one of the three available prompts: "Hello/Good afternoon/Hi". Using Speech Recognition Technology (SRT), the user records his/her answer and begins a conversation with the chatbot. The conversation flows seemingly spontaneously, but when interrupting the series and asking a question (e.g., "How often do you play footbal?"), if the user is not quick enough to pose a question he/she will not get an answer. The reason being that the chatbot leaves insufficient amount of time to ask a question but rather moves on to the next one.

TABLE II. ANALYTICAL EVALUATION OF THE CHATBOTS FOR LEARNING LANGUAGES.

| Chatbot | Speaker | | | | | | | | Technology | | | | | |
|---|---|---|---|---|---|---|---|---|---|---|---|---|---|---|
| | SL | | L | | D | | M | | | DT | | KB | | A |
| | single | multiple | single | multiple | single | multiple | text | voice | multimodal | scripted | statistical | open | closed | open | commercial |
| Mondly https://app.mondly.com/ | - | + | - | + | - | + | + | + | + | + | - | - | + | - | + |
| Andy https://andychatbot.com/ | - | + | - | + | - | + | + | - | - | + | - | - | + | - | + |
| Babbel https://it.babbel.com/ | - | + | - | + | - | + | + | - | - | + | - | - | + | - | + |
| Lanny https://web.eggbun.net/ | - | + | + | - | - | + | + | - | - | + | - | - | + | - | + |

The conversation neither comes across as very authentic (it is obvious that there is a script it follows very closely), nor does it evolve beyond the simple question answering sequence. If the user is an absolute beginner, or even starting over, Mondly is a great choice as it will teach a lot of useful vocabulary and phrases for a beginning language learner. It also works well with lower intermediate students. However, for a more fluent speaker, Mondly is not very challenging. Overall, it can help practice functional speaking skills, but other than that is fairly mundane and unimaginative. It also offers 33 different languages and targets different skills

(speaking, listening, reading, writing). Among other qualities, we would single out personalization, leaderboards, and self-progress reports. The chatbot combines VR technologies as conversations with digital avatars (personas) in artificial environments (such as train, restaurant).

**Andy** is a phone application aimed at helping the user acquire new words/phrases and grammar by practical everyday conversation. The free trial period lasts for 7 days, and if the user wishes to keep practicing after it ends, it charges for more lectures. The Premium subscription gives access to additional learning functions such as grammar lessons and unlimited vocabulary practice for a price. The upside of the application is that the conversation, unlike with Mondly, appears to be spontaneous and extensive. The user almost does not have the impression of talking to a chatbot but rather conversing with a real person. It is quite engaging, and if the user wants to brush up his/her English and get the feel of talking to a real foreigner, Andy is a fairly good option. Speaking of the level, Andy offers greater diversity in that sense. Even though it does not provide a placement test, the impression is that the more fluent users are, the more dynamic and complex the conversional will become and vice versa. However, one of the major issues we had with the chatbot is that it had no voice-enabled function, which leaves the user no other option but to text. This is something that at times may appear as quite a wearisome activity that simply takes too long. Consequently, even despite the chatbot's attempts to keep the user engaged in the conversation, it is inevitable that even the most persevering users will ultimately yield. Another issue is that if the user does not subscribe to either its monthly or yearly package, the chatbot will ask to donate the money quite early on into the conversation, which does strike one as quite odd.

**Babbel** is a virtual assistant that speaks in 14 languages. The chatbot offers dynamic and flexible conversation threads. It supports scaffolded learning by monitoring the user's progress and providing level-based lessons. The lessons are contextual, bit-size lectures (10-15 minutes) covering different topics concerning everyday life situations. The assistant targets a variety of skills, including speaking, listening, reading, and writing. However, it is free only for the first month of use and charges its users after that.

**Lanny** is a live chat app created for people wishing to master the basics or advance their existing knowledge of Korean. One of the benefits of using this chatbot is that, in addition to conversing in Korean, the chatbot also uses English to answer and ask questions and clarify things. This makes Lanny ideal for people trying to grasp the basics of this language. Lanny is more than just a conversational partner, it is also a guide to Korean culture and its history. The application boasts the use of audio by native Korean speakers that will help learners with pronouncing, reading, speaking and writing Korean. The application is available for smartphone users and also has a Web version.

## V. IMPLICATIONS FOR FUTURE WORK

Intelligent chatbots for learning languages are meant to work with professionals, not to replace them. The language experts have lots of knowledge and experience that the most sophisticated Artificial Intelligence (AI) algorithm today cannot match. In these applications and almost all AI applications, it is a major research challenge to study how humans and chatbots can work in synergy to improve the efficiency and effectiveness of technology-supported language learning.

We propose guidelines for future language-learning chatbot design and implementation that bring together language teaching experts, technology designers, and end-users. The reference architecture based on the guidelines identifies and connects critical components of such technology, their purpose and roles. This balanced design and development path can promote chatbots as successful language learning partners.

*Implications for knowledge and education*

We elaborate on three prospective directions as below.

**A new methodology for a technology-supported language learning** that builds on the paradigm of conversational agents. The methodology consists of theoretical, empirical, and technological components such as best practices, training materials, algorithms and programs. The methodology should focus on the following, crucial processes.

*Chatbot's workflow design and implementation* should develop techniques and algorithms for dialog management. The dialogue building approaches can be classified as using retrieval-based models (repository of predefined responses selected by rule-based expression matching) and generative models (creating responses using machine learning techniques and contextual information about the conversation) [27]. The former case may include the structured dialogues developed directly from the reference learning materials. In the latter case, the chatbot automatically learns through conversations [28]. At the moment, robust, but domain-specific dialog systems using handcrafted rules can be created accurately and straightforwardly using available tools such as Google DialogFlow [29] or ChatScript [30] ML tools, such as Google TensorFlow [31], can improve the naturalness of such dialogs. However, the accuracy of the tools ultimately depends on the availability and quality of a language-specific training dataset (or model). The existing datasets are biased towards the English language, while other languages are underrepresented [1]. It makes it more challenging to use ML algorithms to learn languages other than English. On the one hand, the creation of these datasets would require manual effort from proficient speakers. On the other hand, it would embrace the socio-cultural characteristics of the specific language speaking region, which is important when learning a foreign language.

*Chatbot's user interface (UI) design and implementation* should concentrate on its appearances and conversational styles (including gamification elements) to elicit positive attitudes (as the quantity of engagement) and outcomes (as the quality of engagement) [32]. For instance, simulating human languages with high message interactivity (conversational cues). It assumes engaging in back-and-forth message exchanges (mainly textual). Next, the use of human figures (visual cues) leads users to treat chatbots as human and act socially towards them. Finally, human-associated names or identity in the form of labeling a chatbot (identity cue) can be used since humans tend to perceive things by their labels. The form (usability) and function of an interactive product are equally important and cannot be separated. It is important to investigate how to combine the cues to influence user motivations and language proficiency. For example, how to combine them to compensate for the lack of face-to-face communication. It is also relevant to examine the effectiveness of different input modes of the chatbot (speech and text). Implementing such methodology requires cross-cultural, longitudinal qualitative and quantitative studies with end-users. The studies should combine different instruments such as subjective, self-report ratings containing measures such as usefulness, usability, and emotions, and objective metrics such as the level of acquired skills in a foreign language, the amount of conversations, number of errors related to use, and dialogue time. These are all essential predictors of users' technology acceptance and adoption [33].

---

[1] There are 7,000 languages in use around the world - this is why they matter: https://www.weforum.org/agenda/2020/02/how-some-endangered-languages-thrive (Retrieved on 12.07.2020).

**Technology for personalized language learning.** A chatbot which keeps track of the users' past questions and level of language use, could, over a series of interactions, become familiar to users. The chatbot could reuse past language that has been successfully responded to, thereby enhancing users' self-perceived ability as support for interest development. Finally, it could remind the user of the necessity of practice and trying out new words or phrases. We envision defining particular scenarios that match specific language learning tasks and domains using co-design techniques [34].

**A novel, online conversational service for learning languages at scale (L@S)** by creating individual learning experiences, increasing learning outcomes, and supporting lecturers. This approach is especially useful for large-scale lecturers and massive online courses with more than a hundred students per lecture where individual support is not possible due to financial and organizational restrictions. The proposed methodology can deliver a blueprint of the technical infrastructure for developing and deploying such a service.

*Implications for people and society*

We describe how chatbots can support people in their everyday activities and transform our society.

Chatbots present an **affordable and ubiquitous source of language interaction** for many students learning English as a Foreign Language. It is, therefore, essential to understand how they can be put to use best, both inside and outside formal education. As a source of conversation practice and ubiquitous technology, they enable extra practice during independent study anywhere and anytime, inside and outside formal education.

The chatbot can offer some **language learning mechanisms** that many human language learning partners could not and/or would not. In particular, a broader range of expressions/questions and vocabulary, the scaffolded introduction of new vocabulary, grammar, and expressions, or consistent understandable repetition which a human partner is unlikely to present (except language experts). It can also alleviate the issue of availability of native speakers or expert-speaking users, or even difficulty in finding a skilled and available speaking partner online.

**An inclusive design and development strategy for language learning chatbot.** The strategy is not about replacing lecturers with chatbots but creating a co-dependent and intelligent relationship between teacher and chatbot, utilizing their strengths and delivering the best student experience. The strategy should aim to improve chatbots' usability by following the principles of conversational UX design [3].

## VI. THE REFERENCE ARCHITECTURE

Based on the implications, we propose a blueprint of the technological architecture for designing and developing chatbots for learning languages (Figure 1).

The architecture aims to address the issues mentioned above by technology design. The *knowledge layer* contains linguistic concepts and terms necessary for the language content generation. Preferably in different languages, but at least two - a learner's native language and a second language. The existing knowledge sources can be exploited [35]. User-related information includes personal information, learning-related information, and dialog-related information. At the *language layer,* the information for learning languages concerning different language domain(s) and purpose(s) is created from the layer below. The layer's content serves as a basis for generating and maintaining conversations with the users through either interactive lecturing or examinations. The *dialog layer* creates and manages conversation threads. In particular, it accepts and interprets user input, makes the response contents with the specific semantics and syntax in one or several languages (dialog management), and creates the output as a natural language. The *UI layer* implements the agent's conversational appearance towards end-users as either text/visual, voice, or combination.

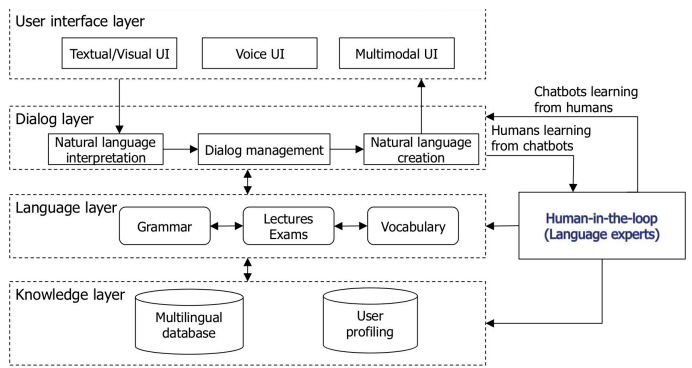

Fig. 1. The reference architecture for conversational agents for learning foreign languages.

Since automatically generated conversations still cannot match the ones created by people, human intervention is necessary. In this mutual relation, humans can learn from chatbots. For example, the new meaning of existing words or new words with new meanings (i.e., casual speech input by users from specific speaking regions or specific dialects). Vice-versa, chatbots can learn from humans through expert supervision. Such supervision may include the manual interventions in the automated natural language interpretation of the learners' input, creation of the responses, and natural language generation. This way, the chatbots can not only improve on their functions but capture the evolution of languages.

## VII. LIMITATIONS

Our goal was to conduct the survey with the analytical (heuristic) evaluation of the chatbots to identify critical aspects, understand their effects, and provide design implications for their improvements. This paper does not provide an exhaustive list of the chatbots for learning foreign languages. Instead, it extracts common examples that describe the current language chatbots' landscape. Besides, we did not assess the effectiveness of the chatbots on learning languages. These are the elements to be covered in future work.

## VIII. CONCLUSION

This paper provides a critical reflection on the current state of the chatbots for learning languages and indicates some important directions for their future. A scarce, resulting sample of the available chatbots (N=4) reveals a complex picture of factors influencing the feasibility of the chatbots for learning languages. Our analysis *informs* end-users of the available chatbot technologies, *communicates* their possibilities to language experts, and *guides* engineers for development. Will the future chatbots for learning languages be fully autonomous, or will the teachers remain as the humans who find out how to augment their practices using the chatbots? For the present, we opt for the latter - at least until AI makes significant advances in natural text and voice recognition, understanding, and generation to equip chatbots with greater effectiveness and autonomy.